\documentclass{scspaperproc}

\usepackage{latexsym}
\usepackage{graphicx}
\usepackage{mathptmx}

\usepackage{amsmath}
\usepackage{amsfonts}
\usepackage{amssymb}
\usepackage{amsbsy}
\usepackage{amsthm}
\usepackage[inline]{enumitem}
\usepackage{tabularx}
\usepackage[disable]{todonotes}
\usepackage{siunitx}
\usepackage{booktabs}
\usepackage{multirow}
\usepackage{caption}
\usepackage{subcaption}
\usepackage{floatrow}
\usepackage[pdftex,colorlinks=true,urlcolor=blue,citecolor=black,anchorcolor=black,linkcolor=black]{hyperref}

\usepackage{hyphenat}
\newtheoremstyle{scsthe}
{8pt}
{8pt}
{\it}
{}
{\bf}
{.}
{.5em}
{}

\theoremstyle{scsthe}

\begin{document}

%
%

\pagestyle{fancyplain}

\thispagestyle{plain}
\firstPageHead{}

\chead{\fancyplain{}{\itshape Sivalingam, Richardson, Tate and Lafferty \vspace{8pt}}}

\rhead{}
\cfoot{}
\renewcommand{\headrulewidth}{0pt} 

\makeatletter
\let\@internalcite\cite
\def\cite{\def\@citeseppen{-1000}%
    \def\@cite##1##2{(##1\if@tempswa , ##2\fi)}%
    \def\citeauthoryear##1##2##3{##1 ##3}\@internalcite}
\def\citeNP{\def\@citeseppen{-1000}%
    \def\@cite##1##2{##1\if@tempswa , ##2\fi}%
    \def\citeauthoryear##1##2##3{##1 ##3}\@internalcite}
\def\citeN{\def\@citeseppen{-1000}%
    \def\@cite##1##2{##1\if@tempswa, ##2)\else{}\fi}%
    \def\citeauthoryear##1##2##3{##1 (##3)}\@citedata}
\def\citeA{\def\@citeseppen{-1000}%
    \def\@cite##1##2{(##1\if@tempswa , ##2\fi)}%
    \def\citeauthoryear##1##2##3{##1}\@internalcite}
\def\citeANP{\def\@citeseppen{-1000}%
    \def\@cite##1##2{##1\if@tempswa , ##2\fi}%
    \def\citeauthoryear##1##2##3{##1}\@internalcite}
\def\shortcite{\def\@citeseppen{-1000}%
    \def\@cite##1##2{(##1\if@tempswa , ##2\fi)}%
    \def\citeauthoryear##1##2##3{##2 ##3}\@internalcite}
\def\shortciteNP{\def\@citeseppen{-1000}%
    \def\@cite##1##2{##1\if@tempswa , ##2\fi}%
    \def\citeauthoryear##1##2##3{##2 ##3}\@internalcite}
\def\shortciteN{\def\@citeseppen{-1000}%
    \def\@cite##1##2{##1\if@tempswa, ##2\else{}\fi}%
    \def\citeauthoryear##1##2##3{##2 (##3)}\@citedata}
\def\shortciteA{\def\@citeseppen{-1000}%
    \def\@cite##1##2{(##1\if@tempswa , ##2\fi)}%
    \def\citeauthoryear##1##2##3{##2}\@internalcite}
\def\shortciteANP{\def\@citeseppen{-1000}%
    \def\@cite##1##2{##1\if@tempswa , ##2\fi}%
    \def\citeauthoryear##1##2##3{##2}\@internalcite}
\def\citeyear{\def\@citeseppen{-1000}%
    \def\@cite##1##2{(##1\if@tempswa , ##2\fi)}%
    \def\citeauthoryear##1##2##3{##3}\@citedata}
\def\citeyearNP{\def\@citeseppen{-1000}%
    \def\@cite##1##2{##1\if@tempswa , ##2\fi}%
    \def\citeauthoryear##1##2##3{##3}\@citedata}
%
%
%
\def\@citedata{%
    \@ifnextchar [{\@tempswatrue\@citedatax}%
                  {\@tempswafalse\@citedatax[]}%
}

\def\@citedatax[#1]#2{%
\if@filesw\immediate\write\@auxout{\string\citation{#2}}\fi%
  \def\@citea{}\@cite{\@for\@citeb:=#2\do%
    {\@citea\def\@citea{, }\@ifundefined
       {b@\@citeb}{{\bf ?}%
       \@warning{Citation `\@citeb' on page \thepage \space undefined}}%
{\csname b@\@citeb\endcsname}}}{#1}}%

%
\def\@citex[#1]#2{%
\if@filesw\immediate\write\@auxout{\string\citation{#2}}\fi%
  \def\@citea{}\@cite{\@for\@citeb:=#2\do%
    {\@citea\def\@citea{, }\@ifundefined
       {b@\@citeb}{{\bf ?}%
       \@warning{Citation `\@citeb' on page \thepage \space undefined}}%
{\csname b@\@citeb\endcsname}}}{#1}}%

%
\def\@biblabel#1{}
\makeatother

\newdimen\bibindent
\bibindent=.25in

\def\thebibliography#1{\section*{\refname}\list
   {}{\settowidth\labelwidth{[#1]}
   \leftmargin \bibindent
   \itemindent -\bibindent
   \listparindent \itemindent
	 \itemsep 4pt
   \parsep 0pt
   \usecounter{enumi}}
   \def\newblock{}
   \sloppy
   \sfcode`\.=1000\relax}

\setlength{\baselineskip}{12.7pt}

\def\SCSconferenceacro{SpringSim}

\def\SCSpublicationyear{2019}

\def\SCSconferencedates{April 29-May 2}

\def\SCSconferencevenue{Tucson, AZ, USA}

\def\SCSsymposiumacro{HPC} 

\title{LASSi: Metric based I/O Analytics for HPC}

\author{
Karthee Sivalingam \\ 
Harvey Richardson \\
Adrian Tate \\ [12pt]
Cray European Research Lab \\
Broad Quay House, Prince Street \\
Bristol, UK \\
\{ksivalinga,harveyr, adrian\}@cray.com\\
\and
\\
Martin Lafferty \\ [12pt]
\\
Cray, UK \\
ACF Building\\
Penicuik, UK \\
rml@cray.com
}
\maketitle
\section*{Abstract}

LASSi is a tool aimed at analyzing application usage and contention caused by use of
shared resources (filesystem or network) in a HPC system. LASSi was initially
developed to support the ARCHER system where there are large variations in application
requirements and occasional user complaints regarding filesystem performance manifested by variation
in job runtimes or poor interactive response. 
LASSi takes an approach of defining derivative \emph{risk} and \emph{ops} metrics that relate to 
unusually high application I/O behaviour.
The metrics are shown to correlate to applications that can experience 
variable performance or that may impact the performance
of other applications.
LASSi uses I/O statistics  over time to provide application I/O profiles
and has been automated to generate daily reports for ARCHER. 
We demonstrate how LASSi provides holistic I/O analysis by monitoring filesystem I/O, generating
coarse profiles of filesystems and application runs and automating analysis of application slowdown using metrics. 

\textbf{Keywords:} I/O, ARCHER, Slowdown, Lustre, Monitoring, Metrics

     \section{Introduction}
\label{sec:intro}

High Performance Computing (HPC) jobs are usually scheduled to run on 
dedicated compute nodes, but will share certain hardware resources with other jobs. In particular, the high-performance interconnect and I/O systems of a supercomputer are typically shared, and so contention can occur when multiple applications/users access these shared resources simultaneously. Shared resources can also be used
inefficiently, for example pathologically bad patterns of communication (affecting the 
network) or inefficient I/O (high metadata rate requirements or small-sized I/O
operations) \cite{opus-nci,nics,nasa}. The combination of these two situations is that poor usage on the part of one user can negatively affect the performance of the shared resource for other users. Users expect consistent 
runtimes but sizing and operating a system to deliver this on an unknown and varied workload 
is very difficult, especially regarding shared resources. In extreme cases user jobs can fail by running unexpectedly past the wallclock time limit requested by the user, resulting in loss of simulation data. Users are reluctant to deal with this by, for example, 
checkpointing. 

LASSi provides HPC system support staff the ability to 
\begin {enumerate*}
[label=\itshape\alph*\upshape)] \item monitor and profile the I/O usage of applications over time  
\item identify and study metrics displaying the quantity and quality of application I/O over time
\item study the risk of slowdown for applications at any time and identify causes for high risk 
\item study rogue applications in detail using profiling tools to identify issues 
at the application level and suggest functional or code changes. \end{enumerate*}
LASSi aims to provide early warning and health status metrics to support staff, enabling much faster triaging of potential I/O issues and the high-level diagnosis of I/O problems. 

\subsection{Background}
The UK's national supercomputing service ARCHER (https://www.archer.ac.uk) supports a highly-varied workload of applications from a range of disciplines including Weather \& Climate, Materials Science, Computational Chemistry, Computational Fluid Dynamics, 
Turbulence research, Quantum Mechanics, High Energy Physics, Biomolecular simulation and Mesoscale engineering along with emerging technologies in AI and Data science.
These applications have different compute and data requirements but
share a common Lustre~\cite{braam2003lsa} file system. This sharing can introduce contention that may impact performance. The severity of the performance impact can be severe enough to affect a user's ability to list directory information.

Users can be quite sensitive to runtime variation or slowdown of submitted jobs. 
Application owners usually submit many similar jobs and 
expect them to complete on time. A \emph{slowdown event} is when a few loosely concurrent jobs run slower than their respective \emph{expected} runtimes. Unfortunately, there is no more precise definition of \emph{expected} runtime than to roughly correspond to the user's wishes. ARCHER support staff have the responsibility to analyse the reasons
for slowdown and then suggest corrective actions. Slowdown can be attributed to many 
factors that also include changes in scientific configuration, node configuration, filesystem load and network
traffic. It has been observed that a few rogue applications may cause slowdown for all users.  
The diverse workload running on ARCHER does not allow a single solution for all such issues. 

ARCHER supports many application that are I/O bound and a detailed study the system's I/O load~\cite{turner_sloan-murphy_henty_richardson} has discussed which file layouts and Lustre striping settings are to be used for optimal
performance and scaling. Many efforts have been made to educate the community through
lectures and training events~\cite{henty_train,epcc_train}. Although these activities are helpful, problems continue to be seen and it is important to focus on problem remediation as well as I/O optimization.

Analysing the slowdown of applications and modeling runtime of jobs in a HPC system is highly complex and time-consuming. Thus, slowdown events incur a high cost to any HPC site or service provider in terms of staff time. LASSi was developed to vastly decrease the amount of time and effort (and cost) required to detect, diagnose and remediate such issues. 

\section{I/O Monitoring and Statistics}\label{sec:monitoring}

LASSi combines Lustre statistics and job information in order to calculate derived metrics. I/O statistics are collected using a bespoke tool called LAPCAT which in turn uses Cerebro~(https://github.com/lmenezes/cerebro) to collect Lustre statistics, storing them in a MySql database on a management server. LAPCAT was developed by Martin Lafferty of Cray UK. Job information is obtained from the job scheduler and ALPS~\cite{cray-alps} logs. On ARCHER, LASSi combines the per-node I/O statistics with the job time information to attribute I/O statistics to individual application launches. The jobstats feature available in newer versions of Lustre can provide some of this information. 

\subsection{ARCHER}
ARCHER is the UK's national supercomputing facility and is a Cray XC30~\cite{cray_xc30}
supercomputer. 
A high-performance Lustre storage system is available to all compute nodes 
and is based on Cray Sonexion 1600 storage running Lustre 2.1.  This storage
system provides 4 filesystems configured from multiple storage units - 
Object Storage Targets (OSTs).  The \emph{fs1} filesystem has 8 OSTs,
\emph{fs2} has 48 OSTs, \emph{fs3} has 48 OSTs and \emph{fs4} has 56 OSTs. 
These filesystems have to support the wide variety of
application domains which produce a complex workload with varying I/O requirements at any given time.

Application runtimes are a function of many factors that include compute clock speed, memory 
bandwidth, I/O bandwidth, network bandwidth and scientific configuration (dataset size or complexity).
Application run time variations due to change in compute resource and memory can be 
ignored. The I/O system and network are shared resources
and are the main causes of slowdown whereas changes to scientific configuration are beyond the scope of LASSi. 

\subsection{Lustre}
Lustre is a distributed parallel filesystem with two important components:
the \emph{Object Storage Server} (OSS) and the \emph{MetaData Server} (MDS). The I/O operation statistics on each server can be 
used to study application I/O usage/performance.
LASSi uses the following I/O statistics:
\begin {enumerate*}
[label=\itshape\alph*\upshape)] 
\item \textbf{OSS:} \emph{read\_kb, read\_ops, write\_kb, write\_ops, other} 
\item \textbf{MDS:} \emph{open, close, mknod, link, unlink, mkdir, rmdir, ren, getattr, 
setattr, getxattr, setxattr, statfs, sync, sdr, cdr}.
\end {enumerate*}

Statistics are aggregated over a time window of three minutes by LAPCAT.
The OSS provides bulk data storage for applications to store data in files. Statistics
\emph{read\_kb} and \emph{write\_kb} refer to the amount of data read and written respectively, while
\emph{read\_ops} and \emph{write\_ops} refer to the number of Lustre operations that are used to achieve 
corresponding read and writes. 
The statistic \emph{other} in OSS refers to the sum of \emph{get\_info,~set\_info\_async,~disconnect,~destroy,~punch,~sync,~preprw} and \emph{commitrw} operations - all relating to the reading and writing of data on the
OSS. The MDS operations relate to filesystem metadata information like
file open and close. The MDS supports creating and deleting objects and controlling application's access to files. Lustre servers provide statistics for both OSS and MDS operations in \emph{stats} files on the filesystem.

\subsection{I/O Statistics}

ARCHER I/O statistics covering a period of 15 months were collected. Initial analysis of the raw statistics revealed great complexity of filesystem usage and individual application I/O profiles. 
LASSi derives higher-level and more practically useful metrics than the raw I/O statistics. At a basic level, the \emph{Relative Standard Deviation} (\emph{RSD}), a common measure of dispersion of a probability distribution, is calculated for each I/O statistic as follows:
\begin{equation}
c_v = \frac{\sigma}{\mu} ,
\end{equation}
where $\sigma$ and $\mu$ are the standard deviation and mean of the data, respectively. Some I/O statistics such as \emph{getxattr,~setxattr,~sdr} and \emph{cdr} are ignored as previous experience shows that they are not prominent. Tables \ref{oss-tab1} and \ref{mds-tab1} show the Lustre statistics of the OSS and MDS respectively for a particular I/O operation that are accumulated per
hour. For example on \emph{fs2}, applications create 105 directories per hour with RSD of 130.
A distribution is considered to be low variance if RSD is less than 1 and so a large RSD value signales a high variance an I/O
statistic.  On ARCHER we generally see a high variance in I/O statistics.
For OSS operational statistics, \emph{fs3} shows very high variance compared to \emph{fs4} and \emph{fs2}. For MDS operational statistics, \emph{fs2} 
shows higher variance than \emph{fs3} and \emph{fs4}.

\emph{fs1} is used for training and we will ignore herein. 
In terms of application hours, \emph{fs3} is used roughly twice as heavily as 
other filesystems. The OSS statistics show a mixed picture, with more reads onto \emph{fs4} and more writes onto \emph{fs3}. Looking at the sum of all MDS operations, \emph{fs4} sees almost twice as many as \emph{fs3} or \emph{fs2}.

\begin{table*}[htbp]
\begin{center}
\caption{OSS Statistics for Lustre filesystems.}
\addtolength{\tabcolsep}{0.05cm}
  \begin{tabular}{l|r|r r |r r|r r|r r |r r}
    \toprule
    \multirow{2}{*}{\textbf{\textit{fs}}} & \multirow{2}{*}{\textbf{\textit{App hours}}} &
      \multicolumn{2}{c}{\textbf{\textit{read\_mb}}}  &
      \multicolumn{2}{c}{\textbf{\textit{read\_ops}}} &
      \multicolumn{2}{c}{\textbf{\textit{write\_mb}}} &
      \multicolumn{2}{c}{\textbf{\textit{write\_ops}}} &
      \multicolumn{2}{c}{\textbf{\textit{other}}}\\
      & {} & {$\mu$} & {$c_v$} & {$\mu$} & {$c_v$} & {$\mu$} & {$c_v$} & {$\mu$} & {$c_v$}& {$\mu$} & {$c_v$}\\
      \midrule
    1	&3447	    &16585 &4 	&150418 &6 	&3783 &7 	&4224 &6 	&313150 &6   \\
	2	&1125513	&5427&13 	&28680&14 	&19904&16 	&26396&14 	&157789&12   \\
	3	&1940595	&4452&26 	&14439&21 	&26187&33 	&33016&28 	&115807&18   \\
	4	&717520	    &13929&5 	&508683&11  &22214&20 	&29367&18 	&1100889&10   \\
    \bottomrule
  \end{tabular}
\label{oss-tab1}
\end{center}
\end{table*}

\begin{table*}[htbp]
\begin{center}
\addtolength{\tabcolsep}{-0.05cm}
\begin{tabular}{l|rr|rr|rr|rr|rr|rr|rr|rr}
\toprule
      \multirow{2}{*}{\textbf{\textit{fs}}} &
      \multicolumn{2}{c}{\textbf{\textit{open}}}  &
      \multicolumn{2}{c}{\textbf{\textit{close}}} &
      \multicolumn{2}{c}{\textbf{\textit{mkdir}}} &
      \multicolumn{2}{c}{\textbf{\textit{rmdir}}} &
      \multicolumn{2}{c}{\textbf{\textit{getattr}}} &
      \multicolumn{2}{c}{\textbf{\textit{setattr}}} &
      \multicolumn{2}{c}{\textbf{\textit{sync}}} &
      \multicolumn{2}{c}{\textbf{\textit{statfs}}}\\	 
& {$\mu$} & {$c_v$} & {$\mu$} & {$c_v$} & {$\mu$} & {$c_v$} & {$\mu$} & {$c_v$}& {$\mu$} & {$c_v$} & {$\mu$} & {$c_v$} & {$\mu$} & {$c_v$}& {$\mu$} & {$c_v$}\\
\midrule
1&45391&9 &45282&9 &  0.8&19     &0.5&31   &1177&18 &541&54  &996&7 & 8 &6 \\ 
2&24314&17 &22040&18  &105&130  &10&67  &13596&10 &6793&14  &317&37 & 1.2 &38 \\ 
3&41547&10 &35389&12  &40&22    &16&29  &13626&14 &1794&22  &23&41 & 5 & 16 \\ 
4&118166&6 &76457&7  &1299&32  &37&17  &20311&16 &2287&14  &32&31  & 3 & 29\\ 
\bottomrule
\end{tabular}
\caption{MDS Statistics for Lustre filesystems.}
\label{mds-tab1}
\end{center}
\end{table*}


Slowdown events are usually reported
to HPC support staff (ARCHER helpdesk) and historically \emph{fs2} has the highest number of such events, with \emph{fs3} seeing the second highest and \emph{fs4} fewer slowdowns. This does not correlate with the combined raw I/O statistics out of LASSi.

\section{LASSi}

LASSi extends the work of Diana Moise \cite{moise_2017} on the Hazel Hen system at the High Performance Computing Center Stuttgart (HLRS), which identified \emph{aggressor} and \emph{victims} 
based on "running at the same time" as an indicator. Grouping applications based on 
the exact command line used, the study defines slowdown as a deviation from the average
run times by 1.5 times or more. This study did not use any I/O or network statistics.

\emph{Victim} detection is based on observing applications that run slower than the average
run time for an application group. \emph{Aggressor} detection is based on applications that overlap with the victims. 
The \emph{aggressor} and \emph{victim} model based on concurrent running becomes difficult to apply when we move to a system like ARCHER, where a large number of applications are usually running. Instead, the LASSi project has defined metrics that indicate problematic \emph{behaviour}. Ultimately, we have shown that there is less distinction between \emph{victim}s and \emph{aggressor} than expected. An alternative explanation, supported by the LASSi derived data is that so-called \emph{victim}s are simply using the Lustre filesystem more heavily than so-called \emph{aggressor}s. 

\subsection{Risk-Metric Based Approach}
We focus on I/O as the most likely cause of application slowdown and
begin with the assumption that in isolation, slowdown only happens 
when an application does more I/O than expected or when an application has an unusually high resource requirement compared to normal. We expect that users will report slowdown only when their applications run at a time when the
filesystem is busier than usual. 

To characterise situations that cause slowdown means considering raw I/O rate, metadata operations and quality (size) of I/O operations. For example, Lustre filesystem usage is optimal when at least 1 MB is read or written 
for each operation (\emph{read\_ops} or 
\emph{write\_ops}). Comparing the \emph{read\_mb}, \emph{write\_mb} with the \emph{read\_ops} and 
\emph{write\_ops} from Table \ref{oss-tab1}, we can infer that the reads are usually sub-optimal 
($\ll$ 1MB) compared to writes.

The central metadata server can sustain a certain rate of metadata operations, above which any metadata request from any application or group of applications will cause slowdown. To provide the type of analysis required, LASSi must comprehend this complex mixture of different applications with widely different read/write patterns, the metadata operations running at the same time and how these interact and affect each other. This requirement informs the LASSi metrics definition.

\subsection{Definition of Metrics}
Metrics for quantity and quality of application I/O
operations must be defined. We first define the risk for any OSS or MDS operation $x$
on a filesystem $fs$ as
\begin{equation}
risk_{fs}(x) = \frac{x-\alpha*avg_{fs}(x)}{\alpha*avg_{fs}(x)}. 
\end{equation}
$\alpha$ is a scaling factor and is set arbitrarily to 2 for this analysis. The risk metric measures
the deviation of Lustre operations from the (scaled) average on a filesystem.
A higher value indicates higher risk of slowdown to a filesystem.

We introduce metrics $risk_{oss}$ and $risk_{mds}$ that accumulate risks to OSS and MDS respectively and are defined by
\begin{equation}
\begin{split}
risk_{oss} &= risk_{read\_kb} + risk_{read\_ops} +  risk_{write\_kb} + risk_{write\_ops} + risk_{other} 
\end{split}
\label{eq:risk_oss}
\end{equation}
and
\begin{equation}
\begin{split} 
risk_{mds} &= risk_{open} + risk_{close} + risk_{getattr} + risk_{setattr} + 
risk_{mkdir}  \\
& +  risk_{rmdir} + risk_{mknod} + risk_{link} + 
risk_{unlink} + risk_{ren}  \\
&+  risk_{getxattr} + risk_{setxattr} + 
risk_{statfs} + risk_{sync} + risk_{cdr} + risk_{sdr} .
\end{split}
\label{eq:risk_mds}
\end{equation}
Non-positive risk contributions are always ignored.

The above metric measures the quantity of I/O operations, but not the quality. On Lustre
1 MB is the optimal size for read or write per operation.
In order to have a measure for the \emph{quality} of application reads and writes we define the metrics
\begin{equation}
read\_kb\_ops = \frac{read\_ops * 1024}{read\_kb} 
\label{eq:read_kb_ops}
\end{equation}
and
\begin{equation}
write\_kb\_ops = \frac{write\_ops * 1024}{write\_kb} .
\label{eq:write_kb_ops}
\end{equation}
The read or write quality is optimal when $read\_kb\_ops =1$ or $write\_kb\_ops=1$.
A value of $read\_kb\_ops >>1$ or $write\_kb\_ops>>1$ denotes poor quality read and writes.
In general, \emph{risk} measures the quantity of I/O and \emph{ops} measures the quality.
 
\subsection{LASSi Architecture}
\begin{figure}[htbp]
\centerline{\includegraphics[scale=0.45]{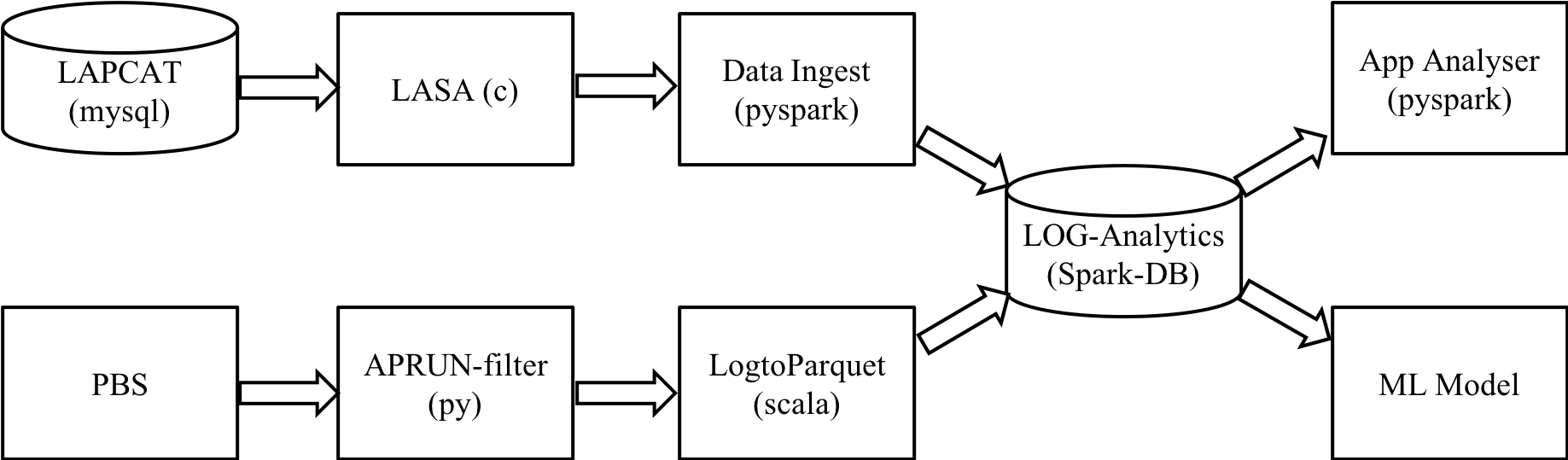}}
\caption{Architecture of LASSi showing different components and 
flow of data through the components.}
\label{fig-lassi-arch}
\end{figure}

LASSi analytics consists of a complex workflow of data movement across different
components developed in PySpark (http://spark.apache.org/docs/2.2.0/api/python/pyspark.html) - a Python API for Spark - C and Scala. 
I/O metrics are computed per application per hour for all three filesystems of ARCHER. They need to be
computed in real-time to enable notification of users or triggering of events in the case of high risk.
Figure \ref{fig-lassi-arch} shows the architecture of LASSi and the data-flow through different components of the tool. 

As noted in Section \ref{sec:monitoring}, the I/O statistics are collected using a tool called LAPCAT at 3-minute granularity. The discrete output may result in errors in I/O statistics attribution at the start and end of application runs. On HPC machines (like ARCHER), applications usually run for many hours and sharp peaks in I/O operations do not affect the application run time compared to sustained high levels in I/O operations. This means that the discretization errors can be easily ignored. Application details including the start time, end time and the compute node list are obtained from the job scheduler. 

LASSi could analyse over 3-minute periods but this might be very expensive.
For practical purposes, LASSi aggregates the data over 60 minutes for analysis. 
All statistics quoted below are using this hourly basis unless mentioned otherwise.
LASA is a C application that aggregates the I/O stats for each application over an hour and 
stores them in a simpler mapping from application ID to I/O statistics for every hour of 
its run. This data is generated in csv format.

Application ID and job ID are not informative but the exact command used to launch the application contains valuable information that can be used to group applications. This grouping was the basis of the victim-aggressor analysis for the initial work \cite{moise_2017}. 
This quantity can be used to find average run times and then study slowdown in application performance. ARCHER uses a PBS scheduler~(https://www.pbsworks.com), ~and APRUN-filter is a python application that filters application information including the exact command in a csv format.

Spark \cite{Zaharia:2016:ASU:3013530.2934664} is used as the data analysis and data mining engine. Spark has an in-built database that supports data import from csv files and also query using SQL.
I/O statistics and job data are stored in relational tables and analysed using SQL queries.
The I/O statistics generated by LASA (in csv format) are ingested by a Spark DB "Data ingest" python tool. The job data is also imported to the Spark-DB using the LogtoParquet \emph{Scala} script. Parquet stores the data in a 
vectorised format that improves the performance of Spark queries.

This data is then aggregated to obtain hourly I/O statistics for all applications running on ARCHER.
The \emph{risk} and \emph{ops} metrics are generated for all application runs every hour by running 
Spark-based SQL queries. The generated \emph{risk} and \emph{ops} profiles are then used for analysis.
LASSi also aggregates statistics for whole groups of applications based on the \emph{run} command used.

The average application run time statistic can be used to study slowdown in application runs. 
This metrics-based framework was developed with the intention of automating analysis on a daily basis, auto-generating plots and reports and potentially providing real-time analysis in the future. Current reporting and plots (see Section \ref{sec:usage}) are generated using python and the 
matplotlib library. 

\section{LASSi Usage and Analysis}\label{sec:usage}
The current LASSi workflow provides
daily analysis of the previous day's filesystem usage. Daily reports generated
by LASSi are accessible to helpdesk and support staff. 
Any slowdown in application run time is usually reported to the helpdesk; the support staff can correlate reported slowdowns of applications to the generated
metrics and identify the application(s) that are causing the problem.  This process of triaging application issues previously consumed significant time and was often inconclusive regarding cause of slowdown. In the case of one Python application that previously caused slow filesystem response, the investigation took several days - similar conclusions can now be reached in a moment using the LASSi tool with automated daily reports. 
\subsection{Daily reports}
\begin{figure}[!h]
    \begin{floatrow}
    \ffigbox{\includegraphics[scale = 0.45]{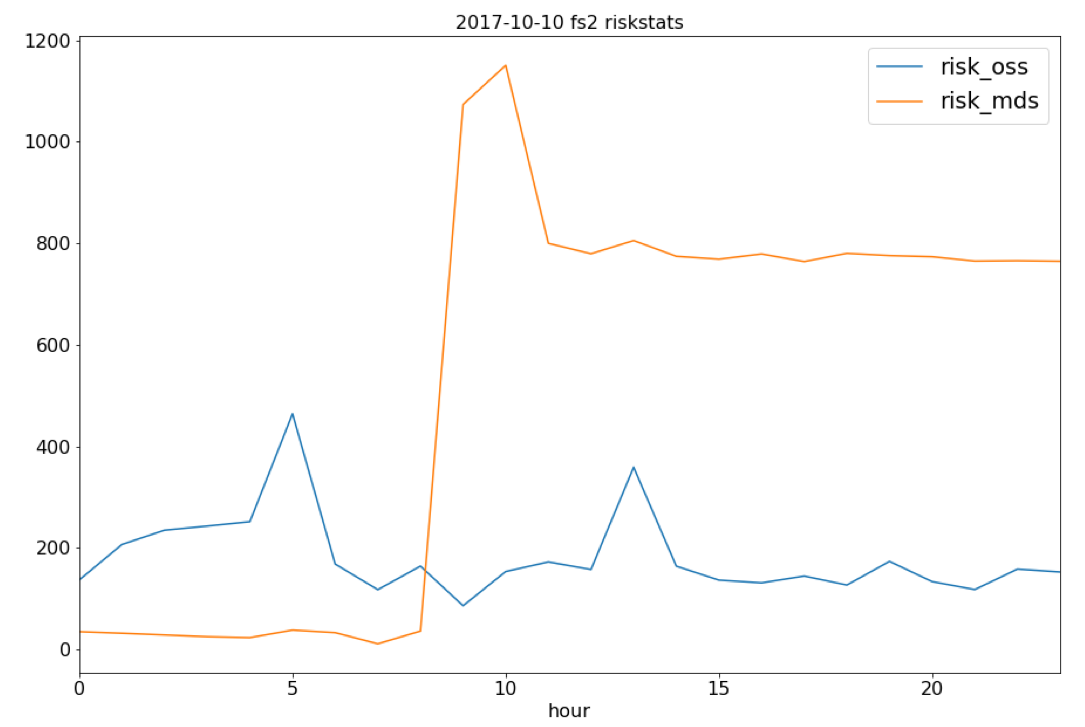}}
    {\caption{Sample report showing the risk (from eqns \ref{eq:risk_oss} and \ref{eq:risk_mds}) to filesystem \emph{fs2} over
24 hours of 2017-10-10.}\label{fig-lassi-risk}}
    \ffigbox{\includegraphics[scale = 0.45]{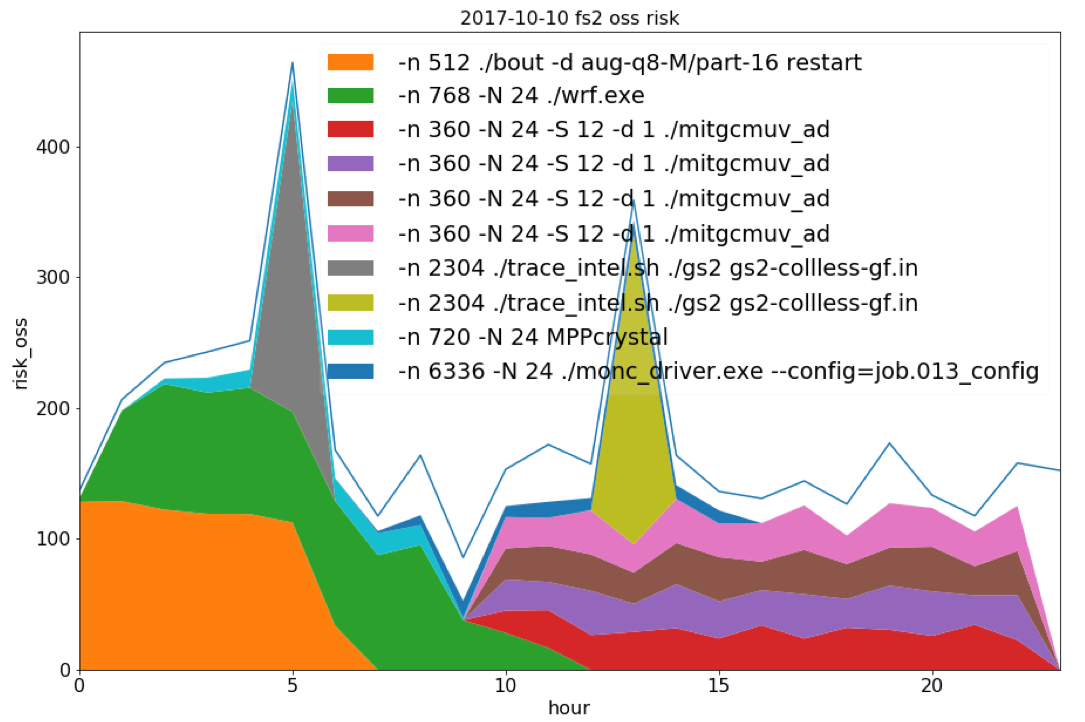}}
    {\caption{Sample report showing the OSS risk to filesystem \emph{fs2} over
24 hours of 2017-10-10 with applications that are contributing to
the risk.}\label{fig-lassi-oss-risk}}
    \end{floatrow}
\end{figure}

LASSi generates daily reports showing I/O statistics and metrics of the previous day for all filesystems. The daily reports contain plots of \emph{risk\_stats}, \emph{mds\_risk},
\emph{oss\_risk} and \emph{ops\_metric}.  LASSi can also generate reports over a specified time period.
The \emph{risk\_stats} plots show the MDS and OSS risk statistics for a filesytem on a certain period.
Figure \ref{fig-lassi-risk} shows a sample report showing  OSS and MDS risk over 24 hours of 2017-10-10 
to \emph{fs2}. These plots can be early indicators of potential slowdown behaviour. 


The \emph{oss\_risk} report shows OSS risk statistics
along with the applications contributing to the risk over time. 
Figure \ref{fig-lassi-oss-risk} shows a sample \emph{oss\_risk} report for filesystem \emph{fs2} on 2017-10-10 and the contributing  applications. Multiple different applications like \emph{bout,~wrf,~mitgcmuv,~gs2,~crystal} and \emph{monc} 
are shown to be causing risk to the filessytem at different times. We see that tracing of \emph{gs2} has peaks in OSS risk, 
while applications like \emph{wrf} and \emph{mitgcmuv} have sustained risk to OSS operations. These reports helped identify multiple cases where slowdown was caused by different applications
running at the same time.

\begin{figure}[!h]
    \begin{floatrow}
    \ffigbox{\includegraphics[scale = 0.45]{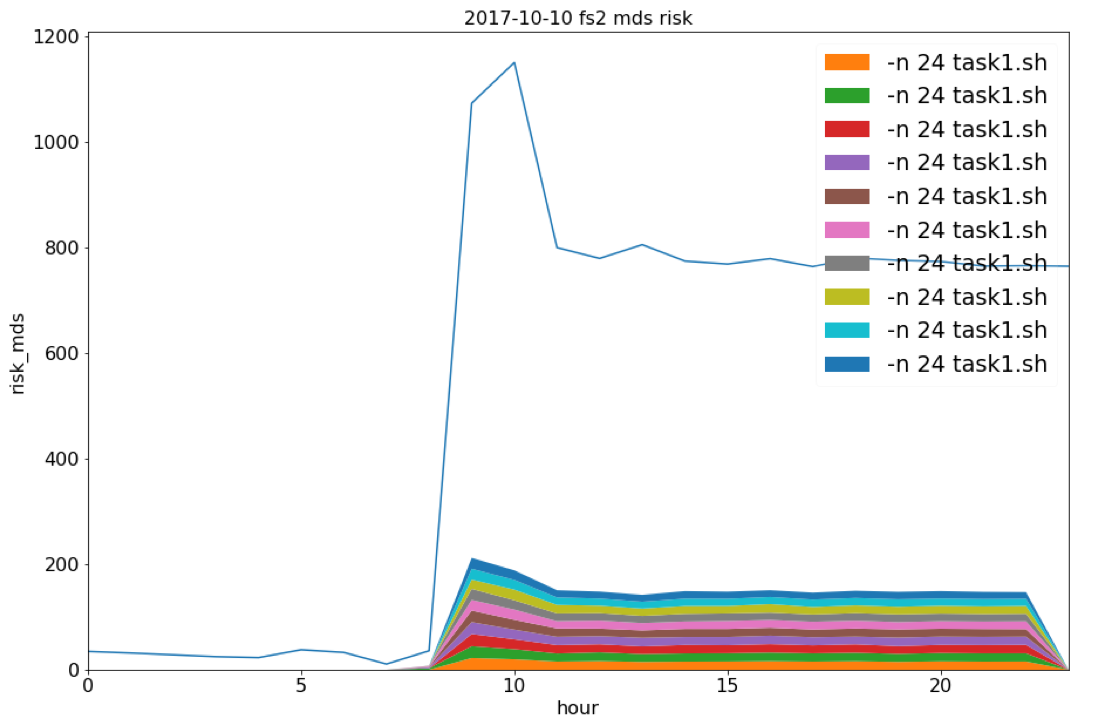}}
    {\caption{Sample report showing the MDS risk to filesystem \emph{fs2} over
24 hours of 2017-10-10 with applications that are contributing to
the risk.}\label{fig-lassi-mds-risk}}
    \ffigbox{\includegraphics[scale = 0.45]{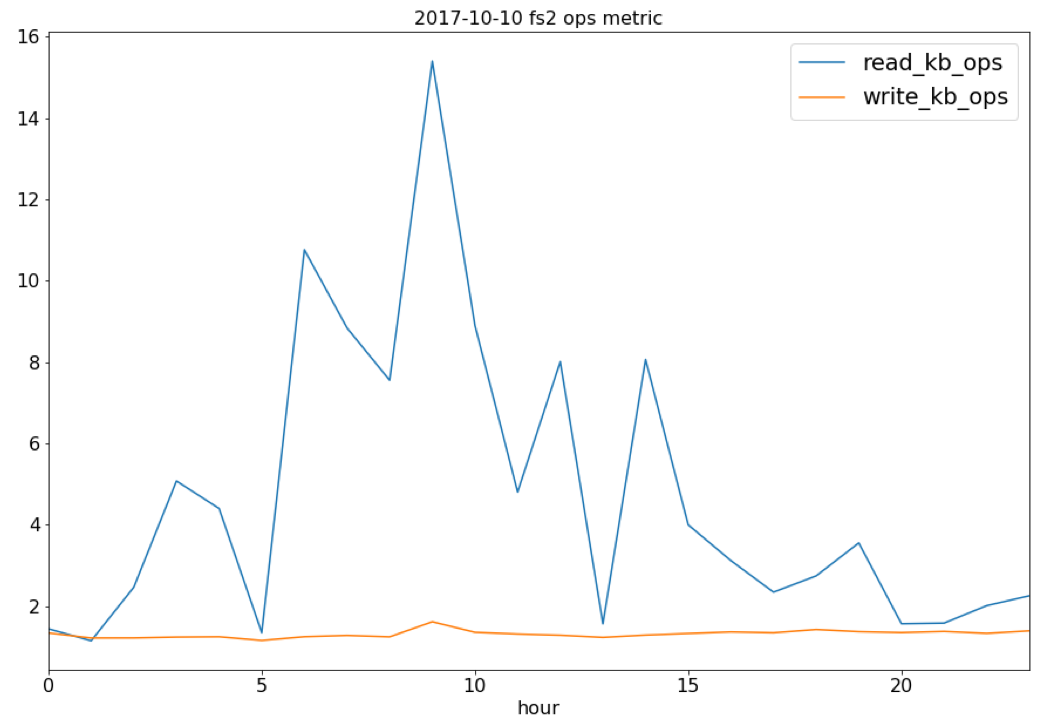}}
    {\caption{Sample report showing the read and write quality (from eqns \ref{eq:read_kb_ops} and \ref{eq:write_kb_ops})
to filesystem \emph{fs2} over 24 hours of 2017-10-10.}\label{fig-lassi-ops}}
    \end{floatrow}
\end{figure}

The \emph{mds\_risk} report shows MDS risk statistics
along with the applications contributing to the risk over time.
Figure \ref{fig-lassi-mds-risk} shows a sample \emph{mds\_risk} report for filesystem \emph{fs2} on 2017-10-10 and the contributing  applications. This is different 
from the \emph{risk\_oss} plot as we see tasks in a taskfarm contributing to the \emph{risk\_mds}. 
Each task contributes to the overall high risk and these are very hard to study and analyse in isolation.  Note that these are not always submitted from a single job or job array. We have already identified a pattern of \lq task farm\rq-like applications
with similar I/O requirements scheduled at the same time causing considerable risk and slowdown.

The \emph{ops\_metric} report shows read and write ops statistics
for a filesystem over time.
Figure \ref{fig-lassi-ops} shows the \emph{read\_kb\_ops} and \emph{write\_kb\_ops} metrics for
\emph{fs2} on 2017-10-10. We observe that the writes are near optimal whereas
the reads are sub-optimal at different time periods. This is a recurring feature in our analysis
as application read quality is usually suboptimal compared to the quality of writes. 


Reports allow HPC
support staff to identify and triage the exact time of risk and the applications that cause risk of slowdown. 
In the case of high OSS risk, attention should be given to the quality of reads and 
writes to ensure that Lustre is optimally used. We observed one tracing application writing a few bytes 
every second to Lustre, which is clearly suboptimal and the problem was resolved by buffering into scratch space.
In case of high MDS risk, the application
should be carefully studied for high metadata operations that contribute to the risk.
One incorrectly configured application was creating millions of directories per second and
this was easily identified using the metrics.
This information is usually passed to the application owner or deep technical support available as part of the ARCHER service who can engage directly with the user. 

In addition to daily monitoring, studying the
metrics of the filesystem helps us understand standard usage of filesystems, define application classes from an I/O perspective and identify
general issues in I/O usage on the system.

\subsection{Application slowdown analysis}
The LASSi \emph{risk} and \emph{ops} metrics we have defined should capture the application slowdown.
Through these metrics and the associated reports, LASSi can
identify application slowdown and assist root cause diagnosis. All
metrics are designed such that higher values are not optimal. Optimal values for \emph{risk}
and \emph{ops} metrics are 0 and 1 respectively. The main contribution factor for slowdown
of an application
is the I/O load (characterised by the metrics) of the filesystem and the I/O profile of that application at any time.
Applications performing no reads and writes will not be impacted by the I/O load
in a filesystem. 

LASSi was partly designed to assist in understanding situations where users report 
performance variation (slowdown) of similar runs. There have been many such incidents reported in 
ARCHER and we have successfully mapped application
slowdown to high risks in filesystems at the time in question. The application(s) causing
high risk are then studied in detail to improve the I/O usage. For reported performance variation, 
we depend on the application owner to clearly label similar job runs and identify slow run times.

\begin{figure}\CenterFloatBoxes
\begin{floatrow}
\ffigbox[\FBwidth]
{\caption{Scatter plot of application run time vs risk of the filesystem for a set of weather/climate jobs.}\label{fig-lassi-val}}
{\includegraphics[scale=0.7]{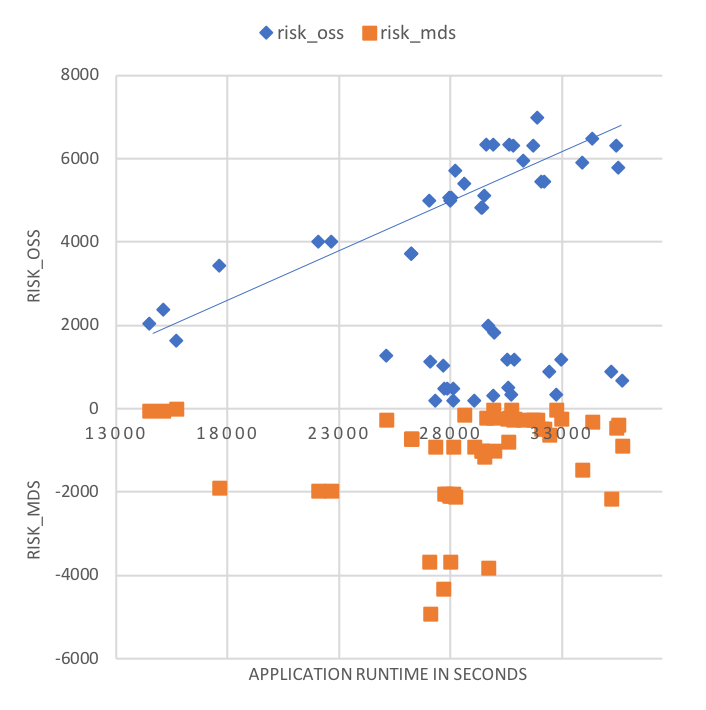}}
\killfloatstyle\ttabbox[\Xhsize]
{\begin{tabular}{||c c c||} 
 \hline
 Job & risk\_oss & risk\_mds  \\ [0.5ex] 
 \hline\hline
job1	&502	&77 \\
job2	&502	&77 \\
job3	&502	&77 \\
job4	&502	&77 \\
job5	&118	&544 \\
job6	&282	&824 \\
job7	&164	&280 \\
job8	&164	&280 \\
 \hline
 \end{tabular}}
{\caption{OSS and MDS risk to filesystem during job runtime.}\label{tab-saturn-jobs}
}
\end{floatrow}
\end{figure}

For example, a user complained about performance variation over 2 days for a Computational Fluid Dynamics (\emph{CFD}) application.
Table \ref{tab-saturn-jobs} shows the sum of risks
to the file system during the job run time. Jobs 1 to 4 ran normally whereas jobs 
5 to 8 ran slowly. The slowdown can be directly mapped to the high metadata risk in the
filesystem during the run times. 
The high risk to OSS does not affect these \emph{CFD} applications. Using LASSi
we can also study the coarse application profile and this \emph{CFD} application was found to 
be doing thousands of meta-data operations (open and close) within each second. 
The high MDS risk to filesystem was caused by \emph{taskfarm} applications running in parallel.
Thus we can map the slowdown to the I/O profile of the application and the I/O load of the filesystem.



Grouping application runs is very difficult and
usually requires the input of the application owner to label the runs that 
are expected to have similar run time. LASSi metrics can be correlated with the run time of application runs, by grouping based on
the exact command used to launch the application. The launch command usually
includes node count, exact node configurations like threads per core, application executable and
application arguments. 

Figure \ref{fig-lassi-val} shows the scatter plot of application run time vs the encountered \emph{risk\_oss} (positive axis) and 
\emph{risk\_mds} (negative axis) in the filesystem for a set of climate and weather jobs.
Here risk metrics are summed over the run time of each application run. The superimposed line in the plot shows a possible linear relationship between \emph{risk\_oss} and run times.
This application group used here has an average 
runtime of 13500 seconds and reads 106MB, writes 14.2 GB and performs 33K metadata operations per hour. 
The average read and write quality are 1.2 and 2.1 and are close to optimal. All these application runs have zero risk with I/O statistics well below the filesystem average.

From the plot, we can see higher OSS and MDS risk on the filesystem when 
jobs with run time more than 13500s were running, with a cluster showing a possible 
linear relationship for \emph{risk\_oss} and 
application run time. The high OSS risk was found to be caused by a python application 
that was reading and writing a few bytes per second at that time.
There is also a cluster of jobs with lesser OSS risk having a run time of more than 23500s which cannot be explained from the risk metrics alone. 
A complete analysis is not possible without understanding the application's science, I/O profile and network bandwidth of each job run. This slowdown analysis did not require the input of the application owner, unlike the previous analysis.

Although LASSi only considers I/O statistics, it has been successful in modeling and resolving slowdown 
incidents reported by application users for over 6 months. In all cases applications
causing slowdown have been identified using \emph{risk} and \emph{ops} metrics and appropriate remedial action had been taken. This approach is more generally applicable to any environment with a shared filesystem as long as the relevant data can be collected.

\section{Related Work}
UMAMI \cite{Lockwood:2017:URG:3149393.3149395} uses an approach of analysing I/O 
statistics using meaningful metrics in a similar fashion to LASSi. They stress a need for a holistic I/O analysis as their metrics do not 
capture enough details to indicate performance loss.
MELT~\cite{DBLP:journals/corr/BrimL15}, a unified Lustre performance monitoring and analysis infrastructure tool,
helps administrators analyse reported application slowdowns by providing command line utilities to
view  I/O statistics of clients, servers and jobs. Using MELT requires expertise and does not provide an automatic root cause analysis solution for performance problems.
ldiskfs \cite{roland_laifer} is a tool for generating Lustre I/O stats for jobs. The script runs hourly
and collects and summarises the jobs I/O stats and then mails the user.
Lustre Monitoring Tool (LMT)~\cite{LMT} is an open-source tool for capture and display of Lustre file 
system activity. I/O statistics are stored in a MySQL database with command
line utilities for live monitoring. LMT does not map I/O statistics to jobs.
Kunkel et al.~\cite{2018arXiv180704985K} review existing tools for analysing I/O performance of parallel 
system and  online monitoring tools developed at DKRZ and LLView by LLNL. They reveal how these tools
can be used to study I/O issues.
Mendez et al. \cite{Mendez:2017:API:3101112.3101260} evaluated I/O performance of applications as a function of
I/O characteristics and performance capacity of the I/O system by defining a metric called 
I/O severity. This metric identifies the factors limiting the I/O performance of a kernel
or application but does not study the effects of multiple applications interacting with the I/O system.
Researchers at NERSC \cite{Uselton2013AFS} introduced a new metric named File System Utilisation (FSU) based on series of calibration experiments using IOR, to study I/O workload on the file system.
Many monitoring tools \cite{Uselton2013AFS}, \cite{Berkeley2009DeployingSF},
\cite{Shipman2010LessonsLI}, 
\cite{uselton2010file}, and
\cite{miller_article}
 for raw I/O statistics of filesystems and jobs have been used to study and improve I/O performance of applications. 
The tools described above provide raw I/O statistics of filesystem or applications. LASSi moves beyond this by delivering a framework where it is easy to identify applications with unusual I/O behaviour, and by targeting application interactions with the filesystem. LASSi is an non-invasive approach that does not perturb the filesystem. 
Additionally, LASSi provides holistic I/O analysis by monitoring filesystem I/O, generating
coarse profiles of filesystems and application runs in time and automating analysis of application
slowdown using metrics. LASSi can also be used to study I/O patterns of application groups which is
important for those that manage filesystems.

\section{Conclusion}
LASSi is a tool primarily designed to help HPC support staff triage and resolve issues of application slowdown
due to contention in a shared filesystem. LASSi uses a metrics-based analysis in which \emph{risk} and \emph{ops} metrics
correlate to the quantity and quality of an application's I/O. The tool's workflow is automated to
produce near real-time analysis of filesystem health and application I/O profiles. Using the metrics
and analysis, LASSi is being used to study the I/O profile of applications, understand common
I/O usage of application groups, locate the reasons for slowdown of similar jobs and to study filesystem usage in general. For example we have identified a particular class of jobs (task farms) that can generate excessive I/O load even though individual applications are not a concern.
This information can be used not only to optimise applications and avoid slowdown 
but also in the planning and configuration of the HPC filesystem for different projects.  We have shown that the application-centric non-invasive approach based on metrics that is used by LASSi is valuable in understanding application I/O behaviour in a shared filesystem.

\section{Future Work}
ARCHER support staff continue to monitor the LASSi metrics against reported application slowdown and contact
application owners of rogue applications to better understand and optimise their I/O. Using these reported incidents, LASSi metrics
are continuously improved and tuned or new metrics added. Currently our analysis uses a coarse time 
resolution of 1 hour, we plan to move to a 6 minute window with hourly analysis of filesystem health.
The ideas from this work can also be 
ready applied for network statistics and this will be explored in the future.

\section*{Acknowledgment}
This work was undertaken by the Cray Centre of Excellence for ARCHER funded by EPSRC. We would like to acknowledge EPSRC, Cray, ARCHER User Support and User Community for their support.

\bibliographystyle{scsproc}
\bibliography{demobib}

\section*{Author Biographies}

\textbf{\uppercase{KARTHEE SIVALINGAM}} is a Research Engineer at the Cray
EMEA Research Lab. He is part of the Cray Center of Excellence for ARCHER that engages with users to allow them to maximise their use of Cray technologies. He has particlar interest in I/O, Workflows, Optimisation, overlap of HPC with Big data and AI \email{ksivalinga@cray.com}.

\textbf{\uppercase{HARVEY RICHARDSON}} is a Senior Research Engineer at the Cray EMEA Research Lab.  He works on EU-funded research projects and the Cray Centre of Excellence for ARCHER.  He has particular interests in computer architecture and performance,
programming models and language standards.

\textbf{\uppercase{ADRIAN TATE}} is Principal Researh Engineer and Director of the Cray
EMEA Research Lab. He is the technical coordinator of the EU Maestro project and is involved in several other EU-funded projects, mostly related to efficient usage of memory hierarchy \email{adrian@cray.com}.

\textbf{\uppercase{MARTIN LAFFERTY}} is a Senior Systems Engineer at the Cray UK Ltd. His work is currently focused around the ARCHER supercomputer based at Edinburgh University with occasional involvement in other global projects. His main interests are computer architecture,  I/O performance, system optimisation, monitoring tools, archival and complex systems firefighting \email{rml@cray.com}.

\newpage

\end{document}